\documentclass[%
 aip,%
 apl,%
 floatfix,
 amsmath,amssymb,
 reprint,%
 numerical,%
]{revtex4-2}

\usepackage{xcolor}
\usepackage{graphicx}%
\usepackage{dcolumn}%
\usepackage{bm} %
\usepackage{verbatim}

\usepackage[utf8]{inputenc}
\usepackage[T1]{fontenc}
\usepackage{mathptmx}
\usepackage{etoolbox}
\usepackage[textsize=scriptsize, colorinlistoftodos, figwidth=8cm]{todonotes}

\usepackage{marginnote} %

\usepackage[separate-uncertainty]{siunitx}

\usepackage{booktabs}
\usepackage{hyperref}

\newcommand{\tabcref}[1]{\hyperref[#1]{Table~\ref*{#1}}}
\newcommand{\figcref}[1]{\hyperref[#1]{Fig.~\ref*{#1}}}
\newcommand{\eqcref}[1]{\hyperref[#1]{Equation~\ref*{#1}}}
\usepackage[above, below]{placeins}

\makeatletter
\def\@email#1#2{%
 \endgroup
 \patchcmd{\titleblock@produce}
  {\frontmatter@RRAPformat}
  {\frontmatter@RRAPformat{\produce@RRAP{*#1\href{mailto:#2}{#2}}}\frontmatter@RRAPformat}
  {}{}
}%
\makeatother

\renewcommand{\vec}{\bm}

\begin{document}

\title[Single-shot capable measurement of the dispersion of surface acoustic waves]{Single-shot capable surface acoustic wave dispersion measurement of a layered plate}

\author{Georg Watzl}
\email{georg.watzl@recendt.at}
\affiliation{Research Center for Non-Destructive Testing GmbH (RECENDT), Altenberger Straße 69, 4040 Linz, Austria}

\author{Stefan Eder}
\affiliation{Research Center for Non-Destructive Testing GmbH (RECENDT), Altenberger Straße 69, 4040 Linz, Austria}

\author{Martin Ryzy}
\affiliation{Research Center for Non-Destructive Testing GmbH (RECENDT), Altenberger Straße 69, 4040 Linz, Austria}

\author{Mike Hettich}
\affiliation{Research Center for Non-Destructive Testing GmbH (RECENDT), Altenberger Straße 69, 4040 Linz, Austria}

\author{Edgar Scherleitner}
\affiliation{Research Center for Non-Destructive Testing GmbH (RECENDT), Altenberger Straße 69, 4040 Linz, Austria}

\author{Martin Schagerl}
\affiliation{Johannes Kepler University Linz, Institute of Structural Lightweight Design, Altenberger Straße 69, 4040 Linz, Austria}

\author{Clemens Grünsteidl}%
\affiliation{Research Center for Non-Destructive Testing GmbH (RECENDT), Altenberger Straße 69, 4040 Linz, Austria}

\date{\today}

\begin{abstract}
    Established techniques for characterizing a layer on a substrate system via surface acoustic wave (SAW) dispersion measurement are often slow due to the need for scanning excitation or detection positions. 
    We present a method for determining discrete points on the SAW mode at equidistant wavenumbers that requires only a single measurement, overcoming these speed limitations. 
    A pulsed laser, shaped into an array of equidistant lines, generates elastic waves on the sample surface. 
    A vibrometer detects the resulting surface displacement next to the line array. 
    The periodic excitation arrangement results in constructive interference of the SAW at wavelengths corresponding to integer fractions of the line spacing. 
    This leads to distinct peaks in the response spectrum, whose higher orders we term spatial harmonics of the SAW. 
    The known line spacing determines the wavelengths, allowing the peak frequencies to be mapped to discrete points on the SAW dispersion curve, effectively sampling the SAW at specific equidistant wavenumbers. 
    By solving an inverse problem, a model can be fit to the experimental data to obtain properties of the layered system. 
    We demonstrate this technique on copper-clad epoxy laminate and roll-cladded aluminum alloy plates and compare our results with dispersion relation scans and micrography. 
    Additionally, we analyze the method's sensitivity to elastic parameters and layer thickness. 
    This approach offers a viable alternative to established SAW scanning methods, particularly in scenarios where the trade-off of lower wavenumber sampling density is acceptable to achieve exceptionally fast measurements while avoiding moving parts. 
\end{abstract}

\maketitle

Layers and layered systems are prevalent structures in industry and everyday life, with an extensive range of applications. 
Some examples relevant to our study include metal coatings to increase corrosion resistance\cite{Budke1999, Guzman2000, Ono2004, Schuster2015, Bischoff2016} or to facilitate metal joining (brazing)\cite{Caron2014, Li2015, Venkateswaran2019, Grunsteidl2021b} to integrated circuit manufacturing\cite{Beyne2003, Nothdurft2019}. 
Layered systems are commonly analyzed by microscopic investigation of the cross-section from slices cut out of the product\cite{Geels2007}. 
However, due to its destructive nature, this method can only be used on a spot-check basis, and the testing procedure is time-intensive. 
Analyzing the dispersion behavior of guided wave modes, especially surface acoustic waves (SAWs), is an effective non-destructive alternative or complementary technique for the characterization of layered materials\cite{Kim1992a, Kim1992b, Schneider1997, Lee2001, Minonzio2010, Salenbien2011, Yeh2012, Moreau2014, Grunsteidl2021b, Diboune2024}. 
By solving an inverse problem, some combinations of layer or substrate parameters like Young's module\cite{Yeh2012} ($E$), Poisson's ratio\cite{Yeh2012} ($\nu$), density\cite{Schneider1997} ($\rho$) or layer thicknesses\cite{Salenbien2011} ($h$) and their -- possibly treated as anisotropic\cite{Kim1992a, Kim1992b} -- equivalents may be determined. 
While the obtainable parameters and their achievable accuracies depend on the specific measurement conditions, data processing methods, material properties of layer and substrate, and how much they differ, SAW dispersion tends to be more sensitive to $E$ and $\rho$ than to $\nu$ and $h$ in many practical scenarios \cite{Schneider1997} (see also \hyperref[sec:fit_params]{Appendix \ref*{sec:fit_params}}). 

Previously demonstrated techniques generally require measuring wave propagation at multiple distances from the source to experimentally obtain the sample's dispersion relation. 
This can be achieved by employing detector arrays\cite{Minonzio2010, Moreau2014}, which are costly and limit distance resolution. 
Alternatively, the detection or excitation position can be scanned\cite{Kim1992a, Kim1992b, Schneider1997, Lee2001, Salenbien2011, Yeh2012, Grunsteidl2021b, Diboune2024}, which, however, limits achievable measurement speeds.

We present a technique for obtaining the surface acoustic wave (SAW) dispersion to a limited extent that does not require scanning or employing detector arrays. 
We demonstrate the method using laser ultrasound (LUS), which has proven to be particularly suitable for guided wave measurements \cite{Schneider1997, Grunsteidl2021b}, owing to its high bandwidth and contact-free operation \cite{Scruby1990, Ces2012}. 
However, the method is not limited to LUS and, depending on the application, other modalities, e.\,g. electromagnetic transducers (EMATs) \cite{Dixon2001, Wilcox2005, Willey2014}, may be used.

\begin{figure}
    \centering
    \includegraphics[scale=1]{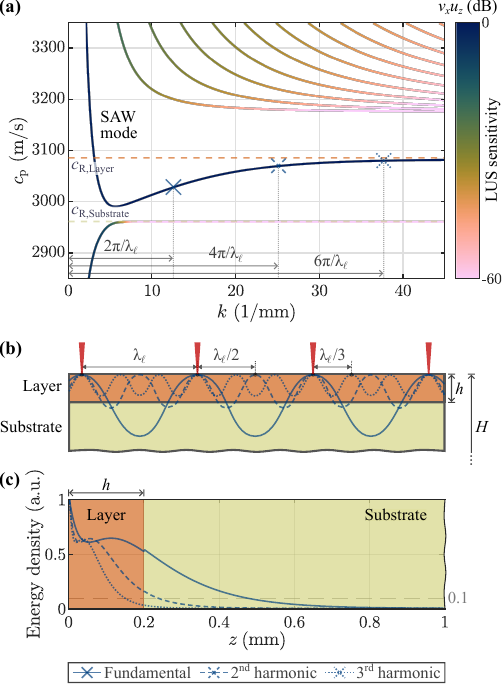}
    \caption[
        Principle of the spatial harmonic SAW method.
    ]
    {
        Principle of the method: 
        (a) calculated dispersion relation of an AlSi alloy layer on an AlMn substrate in $c_\mathrm{p}$ -- $k$ domain; 
        colors indicate the approximate sensitivity of LUS to the modes on the layer's surface;
        (b) illustration of spatial harmonic SAW generation and their propagation depth displayed at the $1/10$ threshold; 
        (c) normalized total (elastic + kinetic) energy densities of the SAW mode points as a function of depth at wavenumbers $k_n$. 
        Calculated using GEWtool\cite{GEWtool} and setting $h=\qty{0.2}{\milli\meter}$, $H=\qty{2.5}{\milli\meter}$, and $\lambda_\mathrm{\ell}=\qty{0.5}{\milli\meter}$.
    }
    \label{fig:harmonic_SAWs} 
\end{figure}

We assume that the investigated samples are laterally homogeneous infinitely extended layered plates. 
Although our approach is, in principle, also suitable for measuring the unidirectional SAW dispersion of anisotropic samples, we restrict this study to the assumption of isotropy for simplicity. 
\figcref{fig:harmonic_SAWs} (a) displays the phase velocity ($c_\mathrm{p}:=2\pi f/k$) dispersion for the guided wave modes of an aluminum-silicon (AlSi) alloy layer with a thickness of \qty{0.2}{\milli\meter} on a \qty{2.3}{\milli\meter} aluminum-manganese (AlMn) substrate with slightly different elastic properties as an example. 
Frequencies are denoted as $f$, wavenumbers as $k$, and wavelengths with $\lambda$. 

In the example shown, the SAW propagating along the coated top surface resides on the second order mode -- except at very low values of $k$. 
Its phase velocity ($c_\mathrm{SAW}$) converges from the Rayleigh (R) wave\cite{Rose2014, Hlavaty2017} velocity of the substrate ($c_\mathrm{R,S}$) towards the R wave speed of the layer ($c_\mathrm{R,L}$). 
This convergence behavior results from the shallower propagation of the SAW mode with increasing $k$\cite{Schneider1997, Glorieux2000, Goossens2007} (see \figcref{fig:harmonic_SAWs} (b)\,--\,(c)). 
Consequently, the SAW interacts less and less with the substrate material with increasing $k$ until it effectively behaves like an R wave propagating in the layer only.

We employ a periodically spaced line excitation pattern with regular line spacings ($\lambda_\mathrm{\ell}$) on the order of a few multiples of the layer thickness ($h$).
This excitation arrangement acts as a spatial comb filter in the wavenumber domain so that constructive interference occurs for guided wave modes with 
\begin{align}
    \lambda_n &= \lambda_\mathrm{\ell}/n \label{eq:lambda_n}   \\
    k_n &= 2n\pi/\lambda_\mathrm{\ell},  \label{eq:k_n}
\end{align}
and $n \in \mathbb{N}_{>0}$.

While modes of other orders are also excited at $k_n$, their amplitudes are negligibly small compared to the SAW when $k$ is sufficiently large\cite{Watzl2025, Diboune2024}. 
This can be inferred from the color-coded overlay in \figcref{fig:harmonic_SAWs} (a). 
It approximates the sensitivity of each point on a mode for a LUS measurement by calculating the LUS excitability (${\propto}\,v_x(z=0)$\cite[chapter 5.3]{Uenishi1998, Scruby1990}) times detectability (${\propto}\,u_z(z=0)$\cite{Monchalin1985}) using the modal surface displacements calculated with GEWtool\cite{GEWtool}. 
Here, $v_x$ refers to the in-plane particle velocity and $u_z$ to the out-of-plane particle displacement at the surface ($z=0$) of the power-flux normalized modal points. 
Therefore, the sensitivity to other modes at $k_n$ is at least one order of magnitude lower so that the resulting response spectrum is dominated by SAW frequency peaks ($f_n$). 

If $\lambda_\mathrm{\ell}$ is known $c_\mathrm{SAW}$ can be calculated at these wavenumbers
\begin{equation}
    \label{eq:c_SAW}
    c_\mathrm{SAW}[k_n] = 2 \pi f_n / k_n\, .
\end{equation}
For $n=1$, this technique is well-established\cite{Cho1996, Maznev1998, Sharples2006, Salenbien2011, Smith2014, Watzl2025} for measuring SAW phase velocities $c_\mathrm{SAW}[k_1] = \lambda_\mathrm{\ell} f_1$. 
The occurrence of higher-order SAW modes ($n>1$) has also been reported previously \cite{Carr1969}, but has not been applied to resolve the dispersion of SAWs propagating in a layer, to the best of our knowledge.

By selecting an appropriate $\lambda_\mathrm{\ell}$ the change of $c_\mathrm{SAW}$ from $c_\mathrm{R,S}$ to $c_\mathrm{R,L}$ can be coarsely sampled.
\figcref{fig:harmonic_SAWs} shows the case with an exemplary line spacing of $\lambda_\mathrm{\ell}=\qty{0.5}{\milli\meter}$ compared to a layer thickness of \qty{0.2}{\milli\meter}.
The fundamental SAW mode (solid blue cross in (a) and solid curve in (b) \& (c)) also propagates to a large extent within the substrate so that $c_\mathrm{SAW}$ is closer to $c_\mathrm{R,S}$ than the higher harmonics.
The $2^\mathrm{nd}$ harmonic's (dashed) energy density decreases by almost $1/10$ before reaching the substrate, and the $3^\mathrm{rd}$ harmonic (dotted) propagates almost exclusively in the layer, leading to $c_\mathrm{SAW}[k_3]\approx c_\mathrm{R,L}$. 
Therefore, the SAW velocity dispersion can be sampled at $k_n$ with a single measurement, and conclusions on the layer can be drawn by comparison to a layered plate guided wave dispersion model\cite{Lowe1995, Hayashi2008, Kiefer2022, Gravenkamp2023}.

\begin{figure}
    \centering
    \includegraphics[width=8.4164cm]{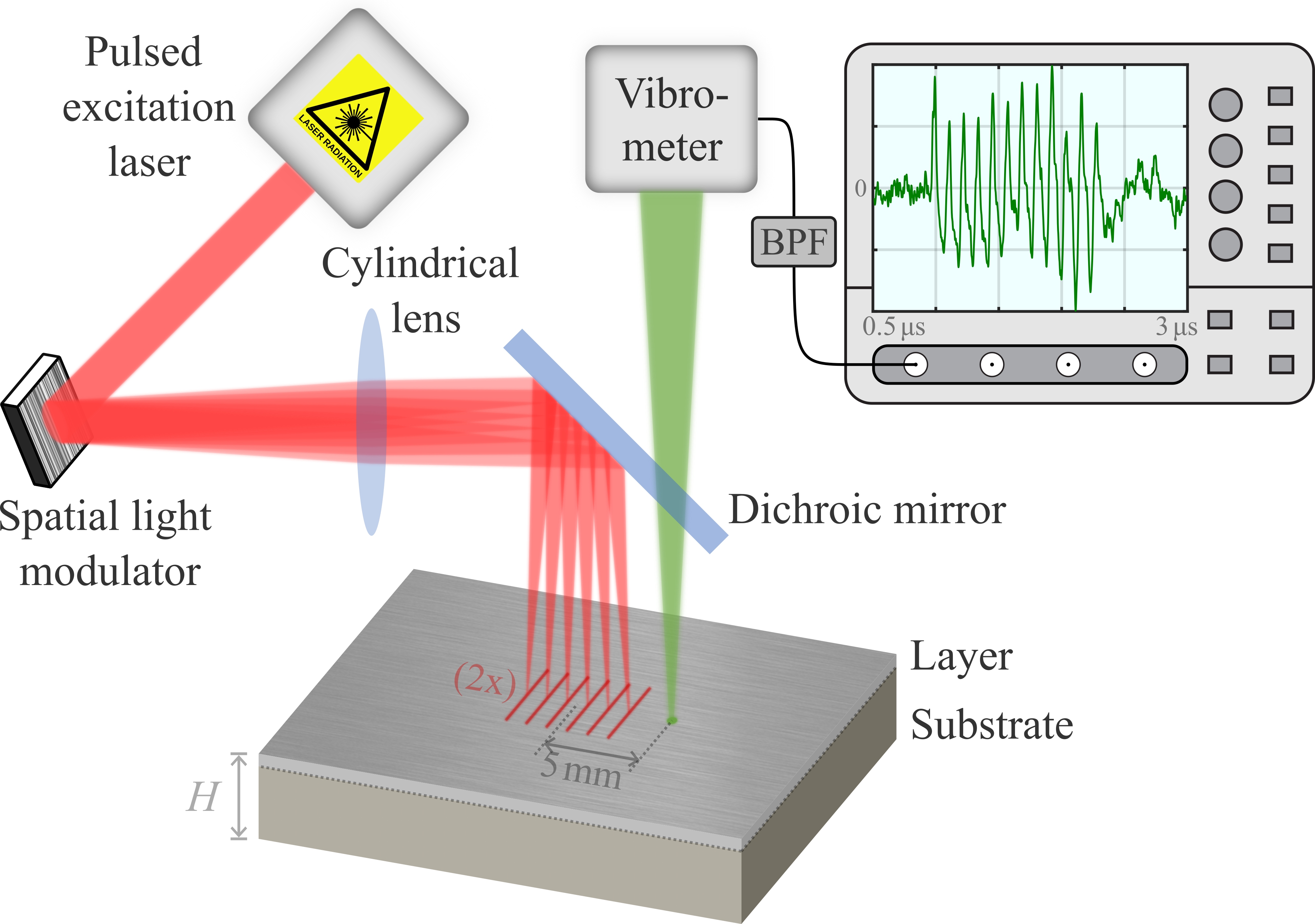}
    \caption{Experimental setup for spatial harmonic SAW measurement with LUS.}
    \label{fig:setup}
\end{figure}

\figcref{fig:setup} displays the experimental setup we developed for single-shot-capable layer characterization. 
The setup is based on the design detailed in our previous study\cite{Watzl2025}, with minor modifications to accommodate the specific requirements of the current investigation. 

For excitation we employed a pulsed laser (Bright Solutions Wedge HB 1064) with a wavelength of \qty{1064}{\nano\meter}, a pulse duration of $\sim$\qty{1.5}{\nano\second}, pulse energies of $\sim$\qty{2}{\milli\joule}, and a pulse repetition rate set to \qty{200}{\hertz}. 
The collimated excitation beam was directed onto a spatial light modulator (SLM) (Holoeye Pluto NIR-149), acting as a diffractive beam splitter. 
This created a periodic pattern of lines in the focal region, which we also analyzed with a beam profiling camera (Cinogy CinCam CMOS-1.001 Nano). 
Each line in the pattern had a full width at half maximum (FWHM) of approximately \qty{40}{\micro\meter} in width and \qty{3.6}{\milli\meter} in length.
The resulting pattern had a line spacing of $\lambda_\mathrm{\ell}=\qty{602}{\micro\meter}$ and was comprised of $n_\mathrm{\ell}=11$ lines. 
These parameters represent a balance between the $k$-domain sampling step size (see \eqcref{eq:k_n}) and spectral resolution\cite{Watzl2025}, as the total pattern size (${\approx}\,n_\mathrm{\ell} \lambda_\mathrm{\ell}$) was constrained by the optical setup. 
For alternative applications of this method, different combinations of $n_\mathrm{\ell}$ and $\lambda_\mathrm{\ell}$ may be adopted depending on the experimental conditions. 
Additionally, the SLM can be replaced by a static diffractive optical element optimized for the specific application.

The detection unit (Sound \& Bright Tempo) is a two-wave mixing interferometer \cite{Monchalin1985, Ing1991} operating at an optical wavelength of \qty{532}{\nano\meter}, that measures $u_z$ at the sample's surface. 
Its detection spot with a size of approximately \qty{50}{\micro\meter} FWHM was positioned adjacent to the excitation pattern along the SAW propagation direction with a distance of \qty{5}{\milli\meter} from the center of the excitation pattern. 
The measured displacement signal was bandpass filtered ($\qty{0.75}{\mega\hertz}<f<\qty{20}{\mega\hertz}$) and digitized with an oscilloscope (Teledyne LeCroy Waverunner HRO 66Zi).

To increase signal-to-noise ratio (SNR), we averaged \num{1000} time traces for each measurement. 
The large areal distribution of the excitation pulse energy allows for much higher pulse energies than our setup could deliver without ablating the sample. 
Combined with commercially available diffractive optical elements with adequate damage thresholds\cite{Katz2018}, this enables a significant reduction in the number of necessary averages, feasibly down to a single-shot measurement.

\begin{figure}
    \centering
    \includegraphics[width=8.5cm]{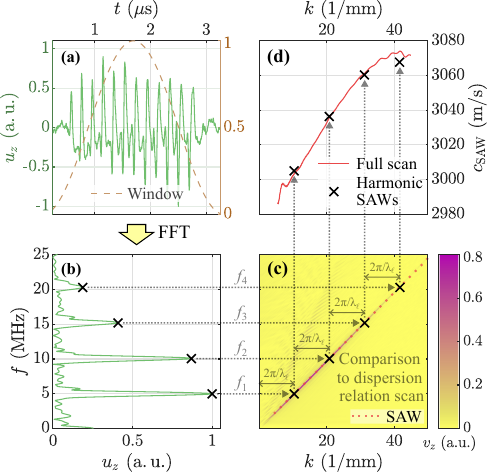}
    \caption[
        Signal processing for the spatial harmonic SAW method. 
    ]
    {
        Signal processing: 
        (a) SAW displacement signal obtained with the setup in \figcref{fig:setup} on a bi-layer aluminum alloy sample ($\approx\qty{0.1}{\milli\meter}$ Al4045 on $\qty{2.4}{\milli\meter}$ Al3003) with $\lambda_\mathrm{\ell}=\qty{602}{\micro\meter}$ and $n_\mathrm{\ell}=12$; 
        (b) Fourier transformed signal with obtained harmonic SAW peak positions; 
        (c) comparison to dispersion map of the same sample obtained from a single-line scan; 
        (d) obtained phase velocities of the SAW mode.
    }
    \label{fig:signal-processing}
\end{figure}

A typical time signal $u_z(t)$ measured with this setup and procedure is shown in the oscilloscope screen in \figcref{fig:setup} and in \figcref{fig:signal-processing} (a). 
Each peak corresponds to the signal emanating from one of the excitation lines. 
Some signal disturbances between peaks can be attributed to higher harmonic SAW propagation, as evidenced by the spectrum in \figcref{fig:signal-processing} (b), where distinct peaks can be identified and their frequencies assigned to $f_n$.
Due to different elastic properties of the layer and the substrate, $f_n$ is not spaced equidistantly (e.\,g., $f_2-f_1 \neq f_3-f_2$).
The wavenumber differences, however, are constant with $\Delta k_n=2\pi/\lambda_\mathrm{\ell}$ (see \eqcref{eq:k_n}).
From $k_n$ and $f_n$, the SAW phase velocities $c_\mathrm{SAW}$ can be calculated using \eqcref{eq:c_SAW}, revealing its dispersive behavior as evidenced by the black crosses in \figcref{fig:signal-processing} (d).

To corroborate our results, we also recorded full dispersion maps ($k$, $f$) of the waveguide by operating the SLM as a mirror and moving the dichroic mirror along the SAW propagation direction and recording $u_z$ at each position\cite{Alleyne1991, Pierce1997}. 
Matching the SAW peaks obtained from a spatial harmonic SAW measurement to the SAW mode on the dispersion map can also be used to calibrate $\lambda_\mathrm{\ell}$ if it is not known or with insufficient accuracy (see \figcref{fig:signal-processing} (c)). 
\figcref{fig:signal-processing} (d) compares $c_\mathrm{SAW}$ extracted from the full dispersion scan\cite{Grunsteidl2021b} with the much faster single-point spatial harmonic SAW measurement. 
With our setup, measuring the full dispersion scan took us about \qty{20}{\minute} of measurement time, using a scanning step size of \qty{50}{\micro\meter}, \num{300} steps and \num{500} averages per position. 
In contrast, the spatial harmonic SAW measurement was finished in about \qty{6}{\second}. 
To obtain mean values and standard deviations, we measured each sample at \num{21} different locations with the spatial harmonic SAW method, increasing total measurement times to about \qty{2}{\minute} per sample. 

\begin{figure}[bht]
    \centering
    \includegraphics{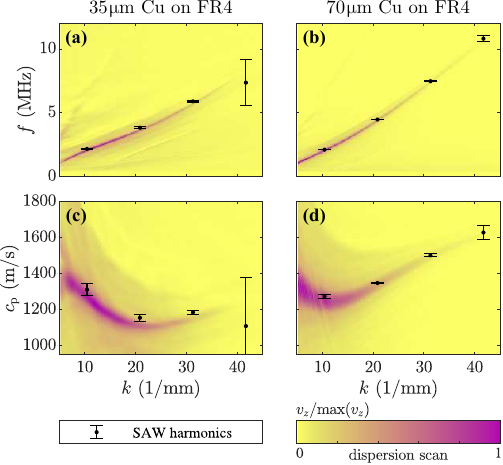}
    \caption[
        Copper-clad epoxy laminate results.
    ]
    {
        Dispersion relation (color map) and spatial harmonic SAW (error bars) measurements on copper-clad fiberglass reinforced epoxy laminate boards with \qty{35}{\micro\meter} (a,\,c) and \qty{70}{\micro\meter} (b,\,d) layer thickness;
        in $f,k$-space (a,\,b) and transferred to $c_\mathrm{p},k$-space (c,\,d).
        The error bars indicate the mean and two times the standard deviation of \num{21} different measurement positions.
    }
    \label{fig:Cu-results}
\end{figure}

\begin{figure*}[bht]
    \centering
    \includegraphics{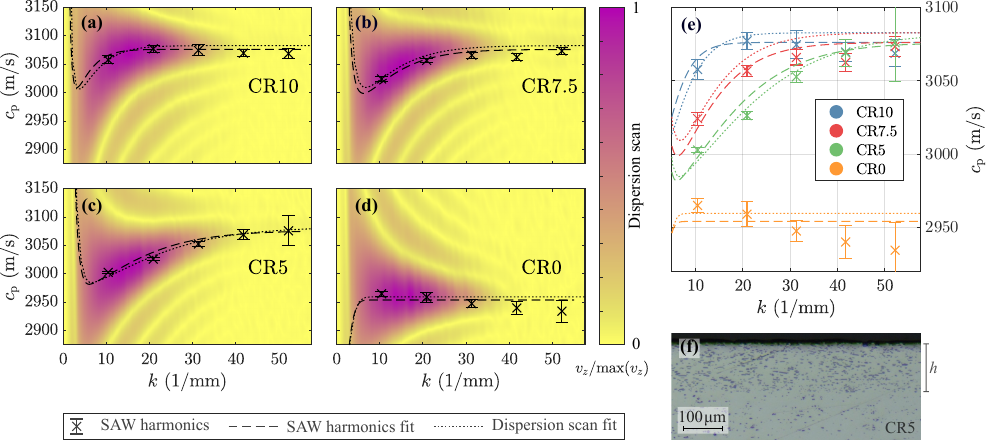}
    \caption[
        Roll-cladded aluminum alloy measurement and model fit results and micrograph image of one of the samples.
    ]
    {
        Roll-cladded aluminum alloy measurement and model fit results (a-e) and micrograph image of one of the samples (f). 
        Error bars indicate two times the standard deviation.
    }
    \label{fig:Al-results}
\end{figure*}

We first demonstrate our method using samples that exhibit a highly dispersive SAW. 
For this purpose, we carried out measurements on copper (Cu) clad fiberglass-reinforced epoxy resin laminates (FR4) with nominal layer thicknesses of \qty{35}{\micro\meter} and \qty{70}{\micro\meter}. 
This sample type is commonly used as a base material for printed circuit boards (PCBs)\cite{Rautio2009} and has very different elastic properties between the layer and the substrate. 
For example, the Young's module of Cu is $E_\mathrm{Cu}\approx\qty{110}{\giga\pascal}$\cite{MatwebCu}, while $E_\mathrm{FR4}\approx\qty{12}{\giga\pascal}$\cite{Kim2007} is an order of magnitude different.

The results displayed in \figcref{fig:Cu-results} show reasonable agreement between the high $u_z$ amplitude regions in the dispersion maps obtained from dispersion relation scans and the spatial harmonic SAW results. 
As expected, a strongly dispersive SAW progression and clear differences between the two layer thicknesses can be observed. 
In contrast, a pure R-wave would be linear in $f,k$-space and constant in $c_\mathrm{p},k$-space. 

However, the achieved SNR is insufficient to resolve the $k$ region where the SAW has converged towards $c_\mathrm{Layer}$ or to consistently resolve more than \num{3} spatial harmonic peaks with reasonable accuracy with the given $\lambda_\mathrm{\ell}$.
This may be due to the high damping of pure copper, compared to many other common engineering metals\cite{Adams1972, Zhang2004}. 
An unknown and potentially varying inhomogeneity and anisotropy of the FR4 substrate between samples\cite{Rautio2009} is a further complication. 
For these reasons, we do not attempt to determine material parameters from these measurements and leave the results as an illustrative qualitative demonstration of our method for substituting the time-intensive scanning approaches that suffer from similar limitations.

As a second example, we demonstrate our method on a layer system with similar layer and base materials and, thus, a rather low SAW dispersion. 
For this purpose, we chose roll-cladded aluminum alloy sheets with different layer thicknesses, used in a previous study\cite{Grunsteidl2021b}, as samples. 
They are comprised of a core of Al3003 and a cladding material of Al4045\cite{Caron2014, Li2015}. 
The measurements designated with CR10 and CR7.5 were performed on samples cladded on both sides, while the CR5 and CR0 measurements were performed on different sides of the same sample, with only one side being coated. 
The samples had a total thickness of $H=\qty{2.5}{\milli\meter}$, except for the CR10 measurement, which was performed on a sample with $H=\qty{4}{\milli\meter}$.
The number in the designation indicates the nominal cladding ratio (CR) on the measurement side relative to $H$.
Therefore, e.\,g. the CR5 measurement has a nominal layer thickness of $h_\mathrm{nom}=\qty{2.5}{\milli\meter}\cdot\qty{5}{\percent} = \qty{125}{\micro\meter}$. 
These nominal thickness values, however, are subject to large manufacturing tolerances of up to \qty{30}{\percent}, so the true values for $h$ at the measurement location can differ significantly from $h_\mathrm{nom}$.

The spatial harmonic SAW results in \autoref{fig:Al-results} coincide well with the high $u_z$ amplitude regions in the dispersion maps obtained by scanning. 
In contrast to the copper-clad FR4 measurements with $c_\mathrm{SAW}$ variations of about \qty{300}{\meter\per\second}, $c_\mathrm{SAW}$ varies in a much smaller range of about \qty{100}{\meter\per\second}.
This time, however, the SNR was sufficient to resolve \num{5} peaks with the same $\lambda_\mathrm{\ell}$ on these samples.
Interestingly, the uncoated measurements (CR0) show minor dispersive behavior in the measurement data instead of remaining constant after converging to $c_\mathrm{R,S}$. 
This could be explained by a microstructural depth gradient due to rolling affecting macroscopic elasticity within the material\cite{Gong2020, Li2022, Watzl2025}, or a grain boundary scattering induced $c_\mathrm{p}$ dispersion\cite{Ryzy2018b, Huang2020}.

We constructed waveguide model representations of the roll-cladded aluminum alloy samples using GEWtool\cite{GEWtool, Kiefer2019} to determine material properties by iteratively minimizing a loss function of model and measurement. 
For the CR10 and CR7.5 measurements, we used a sandwich structure sample model with $h_\mathrm{top}=h_\mathrm{bottom}$ as a constraint, while we used a bi-layer model for the CR5 and CR0 measurements. 
We calculated the full dispersion relation for each optimization iteration step and assigned the SAW to the mode with the highest $u_z(z)$ value at the surface ($z=0$ for CR10, CR7.5, and CR5, and $z=H$ for CR0). 
The sum of squared deviations from the measurement data was used as the minimization value for the spatial harmonic SAW measurements. 
For the dispersion scans, we extracted each value on the dispersion map corresponding to $f$ (or equivalently $c_\mathrm{p}$) and $k$ of the model prediction and took their negative sum as minimization value.

The free parameters were the Young's modulus of the layer ($E_\mathrm{L}$) and substrate material ($E_\mathrm{S}$) as well as the layer thicknesses ($h_\mathrm{CR10}$, $h_\mathrm{CR7.5}$, and $h_\mathrm{CR5}$).
We calculated the densities of layer ($\rho_\mathrm{L}$) and substrate ($\rho_\mathrm{S}$) from the mass fractions of the constituent alloying elements, which were provided by the supplier\cite{Zenodo}, and set them as constant with these calculated values (see \tabcref{tab:Al-fit-results-vert}). 
For Poisson's ratios ($\nu$) we assume values of $\nu_\mathrm{S}=\nu_\mathrm{L}=\num{0.33}$ \cite{MatwebAl4045,MatwebAl3003}. 
The material properties were constrained to be identical for all samples, and the values for $H$ were set as constant with their respective values. 
A discussion on the choice and sensitivity of the fit parameters can be found in \hyperref[sec:fit_params]{Appendix \ref*{sec:fit_params}}. 

The final results of the forward model with the fitting parameters obtained by minimizing with the spatial harmonic SAW measurements or the dispersion scans are shown in \figcref{fig:Al-results} as dashed and dotted lines, respectively. 
While the fits of the scanning and the spatial harmonic SAW measurements do not match exactly, the measurement results and model predictions agree well within each modality. 
Some discrepancies can be seen in the CR0 data measured on the uncoated side, where the model does not predict SAW dispersion in the wavenumber range investigated, as homogeneous layer and substrate materials are assumed.

\begin{table}
    \sisetup{table-alignment-mode=format, table-number-alignment=center, table-format=3.0}
    \centering
    \setlength{\tabcolsep}{5pt}
    \begin{tabular}{cccc}
        \toprule
        \multicolumn{1}{r}{}  & Harmonic SAWs & Dispersion scan  & Reference  \\
        \midrule
        $E_\mathrm{L}$        
            & \qty{77.1}{\giga\pascal}
            & \qty{77.4}{\giga\pascal}   
            & \\
        $\nu_\mathrm{L}$    
            &  
            &  
            & 0.33  \\
        $\rho_\mathrm{L}$   
            &    
            &      
            & \qty{2660}{\kilogram\per\cubic\meter} \\
        $E_\mathrm{S}$      
            & \qty{72.9}{\giga\pascal}  
            & \qty{73.2}{\giga\pascal}   
            &   \\
        $\nu_\mathrm{S}$    
            & 
            &    
            & 0.33\\
        $\rho_\mathrm{S}$ 
            &    
            &       
            & \qty{2729}{\kilogram\per\cubic\meter}  \\
        $h_\mathrm{CR10}$   
            & \qty{504}{\micro\meter} 
            & \qty{369}{\micro\meter}   
            & \qty{362}{\micro\meter} \\ %
        $h_\mathrm{CR7.5}$  
            & \qty{215}{\micro\meter}  
            & \qty{216}{\micro\meter}    
            & \qty{184}{\micro\meter}\\ %
        $h_\mathrm{CR5}$   
            & \qty{149}{\micro\meter}   
            & \qty{122}{\micro\meter}    
            & \qty{116}{\micro\meter}\\ %
        \bottomrule
    \end{tabular}
    \caption[
        Fit results for the roll-cladded aluminum alloy measurements and layer thicknesses from micrography. 
    ]{
        Fit results for the roll-cladded aluminum alloy measurements, reference layer thicknesses from micrography, and reference material parameters set constant for the fit. 
    }
    \label{tab:Al-fit-results-vert}
\end{table}

The resulting material parameters, layer thicknesses, and the parameters set constant for the fit are listed in \tabcref{tab:Al-fit-results-vert}. 
Excellent agreement in $E_\mathrm{L}$ and $E_\mathrm{S}$ between dispersion scan and spatial harmonic SAW results can be observed.
However, the obtained $h$ values, especially of $h_\mathrm{CR10}$, differ significantly. 
As a reference, we also compare the fit results with $h$ obtained from micrographic analysis of cross-sections of three pieces cut from each sample near the LUS measurement position (e.\,g. \autoref{fig:Al-results} (f)). 
These match better with the dispersion scan results, although the $h_\mathrm{CR7.5}$ values similarly deviate by about \qty{15}{\percent} from both dispersion scan and harmonic SAW results. 

The large difference in the $h_\mathrm{CR10}$ results can well be explained by the fact that most of the data points in the spatial harmonic SAW measurements reside within the converged $c_\mathrm{SAW}$ region (see \autoref{fig:Al-results} (a)), whereas more dispersive data is available in the dispersion scan results (see also \hyperref[sec:fit_params]{Appendix \ref*{sec:fit_params}}). 
Another aspect is the much lower sensitivity of $c_\mathrm{SAW}$ to $h$ compared to $E$ for this type of material (see \tabcref{tab:sensitivity}). 
Furthermore, micrographic determination of $h$ is based on the visual distinction of the transition between regions with different silicon precipitate\cite{Lasagni2008} densities. 
As can be seen in \autoref{fig:Al-results} (f), this transition can be fairly blurred, making accurate micrographic thickness readings elaborate and difficult, which further highlights the potential advantages of fast contact-less methods.

In summary, we have presented a contact-free technique for measuring the SAW dispersion of a layer on a substrate without requiring scanning. 
This enables high sampling rates and, given sufficient SNR, can potentially be performed in a single shot, which qualifies the method for moving samples. 
The operating principle of the technique is based on applying a regular periodic excitation pattern on the sample's surface, leading to narrow-band excitation of the SAW at wavelengths matching integer fractions of $\lambda_\mathrm{\ell}$ (\eqcref{eq:lambda_n}). 
As the SAW propagation depth depends on $\lambda$, non-equidistantly spaced $f_n$ peaks can be observed in layered samples when $\lambda_\mathrm{\ell}$ is chosen appropriately (see \figcref{fig:harmonic_SAWs}). 
Thereby, the SAW harmonics contain non-redundant information that makes it possible to determine material properties of the layer and the substrate by solving an inverse problem. 
We demonstrated the technique on roll-cladded aluminum alloy samples and copper-clad epoxy laminate boards, showing the method on samples with high and low SAW dispersion. 
In addition, we analyzed its sensitivity for parameter determination. 
The measurements are consistent with comparative LUS scanning measurements and layer thicknesses obtained from optical micrographs of sample cross sections. 
Although the agreement in $h$ with micrography results is lower than that of the scanning method, this technique can be advantageous in some scenarios due to its high throughput and lack of moving parts.

\hfill

This research was funded in whole or in part by the Austrian Science Fund (FWF) P 33764 and by research subsidies granted by the government of Upper Austria. 
We thank ALRO S.A. for providing the roll-cladded aluminum alloy samples and their chemical compositions. 
For open access purposes, the author has applied a CC BY public copyright license to any author-accepted manuscript version arising from this submission.

\section*{Author declarations}

\subsection*{Conflict of interest}
The authors have no conflicts of interest to disclose.

\section*{Data Availability}

The data that support the findings of this study are openly available in Zenodo at \url{https://doi.org/10.5281/zenodo.14653211}\cite{Zenodo}. 

\appendix

\section{Sensitivity and choice of fit parameters}
\label{sec:fit_params}

The presented method relies on sampling $c_\mathrm{SAW}(k)$ (or equivalently $f_\mathrm{SAW}(k)$) and comparison to a numerical model to obtain quantitative information. 
Consequently the accuracy of determining material parameters and thickness depends on how sensitive $c_\mathrm{SAW}$ reacts to changes in $E_\mathrm{L}$, $\nu_\mathrm{L}$, $\rho_\mathrm{L}$, $E_\mathrm{S}$, $\nu_\mathrm{S}$, $\rho_\mathrm{S}$, $h$, and $H$. 
While an analytical expression for $c_\mathrm{SAW}(k)$ for a layer on a substrate system is not available, it is instructive to analyze its convergence values ($c_\mathrm{R,S}$ to $c_\mathrm{R,L}$) to gauge sensitivity. 

An approximate relation between material parameters and $c_\mathrm{R}$ of a semi-infinite half-space can be derived \cite{Royer2007}
\begin{equation} 
    \label{eq:cR}
    c_\mathrm{R}(E,\rho,\nu) \approx \sqrt{\frac{E}{2\rho(1+\nu)}}  \underbrace{\sqrt{\frac{B(1-\nu)+\nu}{A(1-\nu)+\nu}}}_{g(\nu)}\, ,
\end{equation}
with $A=0.4395$ and $B=0.5797$. 
The function $g(\nu)$ varies only slightly from \numrange{1.05}{1.15} with $0<\nu<0.5$. 
From \eqcref{eq:cR} it follows that the convergence value $c_\mathrm{R,L}$ only depend on $E_\mathrm{L}$, $\nu_\mathrm{L}$, $\rho_\mathrm{L}$, while $c_\mathrm{R,S}$ depends on $E_\mathrm{S}$, $\nu_\mathrm{S}$, and $\rho_\mathrm{S}$. 

Therefore, the convergence values are equally sensitive to $E$ and $\rho$ by a square root relation. 
However, when $E_1/\rho_1 = E_2/\rho_2$, it follows that $c_\mathrm{R}(E_1, \rho_1, \nu) = c_\mathrm{R}(E_2, \rho_2, \nu)$, making $c_\mathrm{R}$ a non-injective function with respect to $E$ and $\rho$. 
This ambiguity makes the simultaneous determination of $E$ and $\rho$ by measuring SAW dispersion difficult. 
Therefore, instead of fitting, we determined $\rho$ from sample weight, geometry, and chemical composition (see supplementary data\cite{Zenodo}). 

On the other hand $\nu$ relates to $c_\mathrm{R}$ by $(1+\nu)^{-1/2} g(\nu)$, which is a much weaker relation than with $E$ and $\rho$. 
We conclude that dispersion curves of layer-on-plate systems are typically much more sensitive to $E$ and $\rho$ than to $\nu$, which agrees with previous analyses found in literature\cite{Schneider1997}. 
However, as with $E$ and $\rho$, numerical relations between $E$ and $\nu$, and $\rho$ and $\nu$ can be found, which yield equal $c_\mathrm{R}$, and the limitations for simultaneous determination of material parameters also applies to $\nu$. 
Since the influence of $\nu$ on $c_\mathrm{SAW}$ is small compared to $E$ and $\rho$ anyway, we did not include $\nu$ in our fit procedure and chose values from literature instead.

The total thickness $H$ only affects $c_\mathrm{SAW}$ for very small values of $k$, which is why we do not attempt to determine $H$ with our method. 
Due to the decrease in propagation depth of the SAW with increasing $k$ (see \figcref{fig:harmonic_SAWs}), $h$ mainly influences convergence speed. 
The sensitivity of $c_\mathrm{SAW}$ to $h$ must be obtained numerically on a case-by-case basis, depending on the materials and $k$.

\begin{table}
    \setlength{\tabcolsep}{1.5pt}
    \centering
    \begin{tabular}{c c S[table-format=-1.2] c S[table-format=-1.2]}
        \toprule
            {\scriptsize{$k_0=\qty{20}{\per\milli\meter}$}}
            & \multicolumn{2}{c}{AlSi-AlMn}    
            & \multicolumn{2}{c}{Cu-FR4}    \\
            \cmidrule(lr){2-3} \cmidrule(lr){4-5}
            {$\phi$}
            &  \multicolumn{1}{c}{$\vec{\Phi}_0$} 
            &  {$\left.\frac{\partial c_\mathrm{SAW}}{\partial \phi}
                \frac{\phi}{c_\mathrm{SAW}}\right|_{\vec{\Phi}_0}$} 
            &  \multicolumn{1}{c}{$\vec{\Phi}_0$} 
            &  {$\left.\frac{\partial c_\mathrm{SAW}}{\partial \phi}
                \frac{\phi}{c_\mathrm{SAW}}\right|_{\vec{\Phi}_0}$}  
         \\ \midrule
            {$E_\mathrm{L}$} 
            &   \qty{77.4}{\giga\pascal}
            &   0.38
            &   \qty{110}{\giga\pascal}
            &   0.31    
        \\
            {$\nu_\mathrm{L}$} 
            &   0.33
            &   -0.05
            &   0.34
            &   0.02    
        \\
            {$\rho_\mathrm{L}$} 
            &   \qty{2660}{\kilogram\per\cubic\meter}
            &   -0.42
            &   \qty{8960}{\kilogram\per\cubic\meter}
            &   -0.43 
         \\ 
            {$E_\mathrm{S}$} 
            &   \qty{73.2}{\giga\pascal}
            &   0.12
            &   \qty{12}{\giga\pascal}
            &   0.19    
        \\ 
            {$\nu_\mathrm{S}$} 
            &   0.33
            &   -0.03
            &   0.13
            &   -0.02    
        \\ 
            {$\rho_\mathrm{S}$} 
            &   \qty{2729}{\kilogram\per\cubic\meter}
            &   -0.08
            &   \qty{1929}{\kilogram\per\cubic\meter}
            &   -0.07  
        \\ 
            {$h$} 
            &   \qty{200}{\micro\meter}
            &   0.02
            &   \qty{70}{\micro\meter}
            &   0.28
        \\ \bottomrule
    \end{tabular}
    \caption[
        Derivatives of $c_\mathrm{SAW}$ for two different layer-on-plate systems.
    ]
    {
        Normalized numerical derivatives of $c_\mathrm{SAW}$ at $k_0=\qty{20}{\milli\meter^{-1}}$ with the parameter set $\vec{\Phi} = \{E_\mathrm{L}, \nu_\mathrm{L}, \rho_\mathrm{L}, E_\mathrm{S}, \nu_\mathrm{S}, \rho_\mathrm{S}, h\}$ and differentiation variable $\phi \in \vec{\Phi}$ around $\vec{\Phi}_0$, whose values represent a $H=\qty{2.5}{\milli\meter}$ thick roll-cladded aluminum alloy plate (AlMn-AlSi, \tabcref{tab:Al-fit-results-vert}) and a copper-clad epoxy laminate board (Cu-FR4) with $H=\qty{1.5}{\milli\meter}$ (SAW dispersion calculated with GEWtool\cite{GEWtool}).
    }
    \label{tab:sensitivity}
\end{table}

An analysis with two sets of parameters that approximate the different sample types (roll-cladded aluminum alloy and copper-clad FR4 boards) used in the demonstration experiments at $k_0=\qty{20}{\milli\meter^{-1}}$ can be found in \tabcref{tab:sensitivity}. 
Corresponding graphical representations covering a larger range of $k$ are available in the supplementary data\cite{Zenodo}. 
It should be noted that the sensitivity analysis does not account for any experimental difficulties such as low SNR or high material damping. 

The results corroborate the previous discussion on \eqcref{eq:cR}: 
The absolute values of the normalized derivatives of $c_\mathrm{SAW}$ are largest for $E$ and $\rho$ (with similar values, but opposing signs) and much lower for $\nu$ within each layer and substrate material. 
Notably, the sensitivity to $h$ in the roll-cladded aluminum alloy parameter set is about \num{10} times lower than to $E$, which may contribute to the differing results for $h$ in \tabcref{tab:Al-fit-results-vert}. 
The sensitivities for the copper-clad FR4 laminate parameter set are similar, except for $h$, where it is an order of magnitude larger. 
This indicates that the accuracy of coating thickness determination by measuring the SAW dispersion benefits most from highly divergent coating and substrate materials.

\bibliography{refs}%

\end{document}